\documentstyle[prl,aps,epsf,floats]{revtex}

\begin{document}

\twocolumn[\hsize\textwidth\columnwidth\hsize\csname@twocolumnfalse%
\endcsname

\title{Long range correlations in the non-equilibrium quantum
  relaxation of a spin chain}

\author{Ferenc Igl\'oi$^{1,2}$ and Heiko Rieger$^{3}$}

\address{
$^1$ Research Institute for Solid State Physics and Optics, 
H-1525 Budapest, P.O.Box 49, Hungary\\
$^2$ Institute for Theoretical Physics,
Szeged University, H-6720 Szeged, Hungary\\
$^3$ Theoretische Physik, Universit\"at des Saarlandes, 
     66041 Saarbr\"ucken, Germany\\
}

\date{January 13, 2000}

\maketitle

\begin{abstract}
  We consider the non-stationary quantum relaxation of the Ising spin
  chain in a transverse field of strength $h$. Starting from a
  homogeneously magnetized initial state the system approaches a
  stationary state by a process possessing quasi long range
  correlations in time and space, independent of the value of $h$. In
  particular the system exhibits aging (or lack of time translational
  invariance on intermediate time scales) although no indications of
  coarsening are present.
\end{abstract}

]

\newcommand{\bc}{\begin{center}}
\newcommand{\ec}{\end{center}}
\newcommand{\be}{\begin{equation}}
\newcommand{\ee}{\end{equation}}
\newcommand{\ba}{\begin{array}}
\newcommand{\ea}{\end{array}}
\newcommand{\beqn}{\begin{eqnarray}}
\newcommand{\eeqn}{\end{eqnarray}}

\newcommand{\psio}{\vert\psi_0\rangle}
\newcommand{\psioa}{\langle\psi_0\vert}
\newcommand{\bra}[1]{\langle#1\vert}
\newcommand{\ket}[1]{\vert#1\rangle}

Non-equilibrium dynamical properties of quantum systems have been of
interest recently, experimentally and theoretically. Measurements on
magnetic relaxation at low-temperatures show deviations from the
classical exponential decay\cite{exp}, which was explained by the
effect of quantum tunneling. On the theoretical side, among others,
integrable\cite{int} and non-integrable models\cite{nonint} were
studied in the presence of energy or magnetic currents, as well as the
phenomena of quantum aging in systems with long-range\cite{long} and
short-range interactions\cite{short}.

Here we pose a different question: Consider a quantum mechanical
interacting many body system described by a Hamilton operator
$\hat{H}$ without any coupling to an external bath, which means that
the system is closed. Suppose the system is prepared in a specific
state $\psio$ at time $t=0$, which is {\it not} an eigenstate of the
Hamiltonian $\hat{H}$. Then we are interested in the natural quantum
dynamical evolution of this state which is described by the
Schr\"odinger equation and is formally given by
\be
\vert\psi(t)\rangle = \exp\left(-\frac{i}{\hbar}\hat{H}t\right)\psio\;.
\ee
Obviously the energy $E=\psioa\hat{H}\psio$ is conserved.
In particular we want to study the time evolution of the expectation
value $A(t)$ of an observable $\hat{A}$ or the two-time correlation
function $C_{AB}(t_1,t_2)$ of two observables $\hat{A}$ and $\hat{B}$,
defined by
\beqn
A(t)            & = & \psioa\hat{A}_H(t)\psio\label{obs}\\
C_{AB}(t_1,t_2) & = & \psioa\{\hat{A}_H(t_1)\hat{B}_H(t_2)\}_S\psio\;,
\nonumber
\eeqn
where $\hat{A}_H(t)=\exp(+i\hat{H}t)\hat{A}\exp(-i\hat{H}t)$ is the
operator $\hat{A}$ in the Heisenberg picture (with $\hbar$ set to
unity) and $\{\hat{A}\hat{B}\}_S=1/2(\hat{A}\hat{B}+\hat{B}\hat{A})$
the symmetric product of two operators.

One should emphasize that in such a situation one does {\it not}
expect time translational invariance to hold, which would manifest
itself in, for instance, $A(t)=A_0={\rm const.}$ and
$C_{AB}(t_1,t_2)=C_{AB}(t_1-t_2)$. There will be a transient regime in
which these relations are violated and depending on the complexity of
the system this {\it non-equilibrium} regime will extend over the
whole time axis, in which we would denote it as {\it quantum aging},
as it can be observed for instance for the universe, which is (most
probably) a closed system.

To be concrete we consider the prototype of an interacting quantum
systems, the Ising model in a transverse field (TIM) in one dimension
defined by the Hamiltonian:
\be
H=-{1 \over 2} \left[\sum_{l=1}^{L-1} \sigma_l^x\sigma_{l+1}^x+
h \sum_{l=1}^L \sigma_l^z \right]\;,
\label{hamilton}
\ee
where $\sigma_l^{x,z}$ are spin-$1/2$ operators on site $l$. We
consider initial many body states that are eigenstates either of all
local $\sigma_l^x$ or of all local $\sigma_l^z$ operators. We will
mostly be concerned with fully magnetized initial states, either in
the $x$ or the $z$ direction, which we denote with $\ket{x}$ and
$\ket{z}$, respectively, and which obey $\sigma_l^x\ket{x}=+\ket{x}$
and $\sigma_l^z\ket{z}=+\ket{z}$, respectively.

In passing we note that one obtains the zero temperature {\it
  equilibrium} situation by choosing the ground state of the
Hamiltonian (\ref{hamilton}) as the initial state. This ground state
has a quantum phase transition at $h=1$ from a paramagnetic ($h>1$) to
a ferromagnetic ($h<1$) phase, the latter being indicated by long
range order in the $x$-component, i.e.\ a non-vanishing expectation
value of $\sigma^x$. Moreover, non-zero temperature {\it equilibrium}
relaxation has been considered in \cite{sy}.

The expectation values and correlation functions we are interested in
are those that originate from these spin operators $\sigma_l^x$ and
$\sigma_l^z$. In order to compute them, we have to express the
Hamiltonian (\ref{hamilton}) in terms of fermion creation
(annihilation) operators \cite{pfeuty,lsm} $\eta_q^+$ ($\eta_q$) 
\be
H=\sum_q \epsilon_q \left( \eta_q^+ \eta_q -{1 \over 2} \right)\;.
\label{freefermion}
\ee
The energy of modes, $\epsilon_q$, $q=1,2,\dots,L$ are given by the
solution of the following set of equations
\begin{eqnarray}
\epsilon_q\Psi_q(l)&=&-h\Phi_q(l)-\Phi_q(l+1)\; ,\nonumber\\
\epsilon_q\Phi_q(l)&=&-\Psi_q(l-1)-h\Psi_q(l)\; ,
\label{I.3}
\end{eqnarray}
and we use free boundary conditions which implies for the components
$\Phi_q(L+1)=\Psi_q(0)=0$.  The spin-operators can then be expressed
by the fermion operators as
\beqn
\sigma_l^x&=&A_1 B_1 A_2 B_2 \dots A_{l-1} B_{l-1} A_l\; ,\nonumber\\
\sigma_l^z&=&-A_l B_l\;,
\label{operators}
\eeqn
with
\be
\ba{lcl}
A_i&=&\sum_{q=1}^L \Phi_q(i)(\eta_q^+ +\eta_q)\;,\\
B_i&=&\sum_{q=1}^L \Psi_q(i)(\eta_q^+ -\eta_q)\;,
\ea
\label{ab}
\ee
and the time-evaluation of the spin operators follows from that of the
fermion operators: $\eta_q^+(t)=e^{it\epsilon_q} \eta_q^+$ and
$\eta_q(t)=e^{-it\epsilon_q} \eta_q$.

To calculate different non-equilibrium correlation functions we have
developed a systematic method\cite{unp} in which the time-dependent
contractions are defined by:
\beqn
\langle A_l A_k \rangle_t&=&\sum_q \cos( \epsilon_q t) \Phi_q(l) \Phi_q(k)\; ,\nonumber\\
\langle A_l B_k \rangle_t&=&\langle B_k A_l \rangle_t=i \sum_q \sin (\epsilon_q t) \Phi_q(l) \Psi_q(k)\; ,\nonumber\\
\langle B_l B_k \rangle_t&=&\sum_q \cos (\epsilon_q t) \Psi_q(l) \Psi_q(k)\; .
\label{contr}
\eeqn
play a central role. For general position of the spin, $l=O(L/2)$, one
finds simple formulas for the expectation values and correlation
functions involving $\sigma_l^z$ operators, whereas the calculation of
those involving $\sigma_l^x$ operators is a difficult task and the
final result is complicated \cite{pfaff}.  However, both the
surface-spin auto-correlations and the end-to-end correlations are
given in quite simple form, both for the equilibrium \cite{lsm} and
for the non-equilibrium case.

First we study the $x$-end-to-end correlations defined by
\be
C^{x,\psi}_L(t)
=\bra{\psi_0}\{\sigma_1^x(t)\sigma_L^x(t)\}_S\ket{\psi_0}\;,
\ee
which contain information about the existence or absence of magnetic
order in the $x$-direction. The single time $t$ at which the
correlations between the two spins are measured indicates the age of
the system after preparation. For the fully ordered initial state
$\ket{\psi_0}=\ket{x}$ we obtain
\be
C^{x,x}_L(t)=
\langle A_1 A_1 \rangle_t \langle B_L B_L \rangle_t 
+ |\langle A_1 B_L \rangle_t|^2\;,
\label{Cxx}
\ee
The first term in the r.h.s. of Eq.(\ref{Cxx}) is the product of surface
magnetizations at the two ends of the chain. Therefore $\lim_{L,t \to \infty}
C^{x,x}_L(t)=\overline{m_1}^2$ and the stationary state, starting with
$\ket{x}$, has long-range order for $h<1$ as $\overline{m_1}=1-h^2$. Thus the
surface order-parameter, $\overline{m_1}$, vanishes continuously at the
transition point, $h_c=1$, with a non-equilibrium exponent, $\beta_1^{ne}=1$.

\begin{figure}[t]
\epsfxsize=\columnwidth\epsfbox{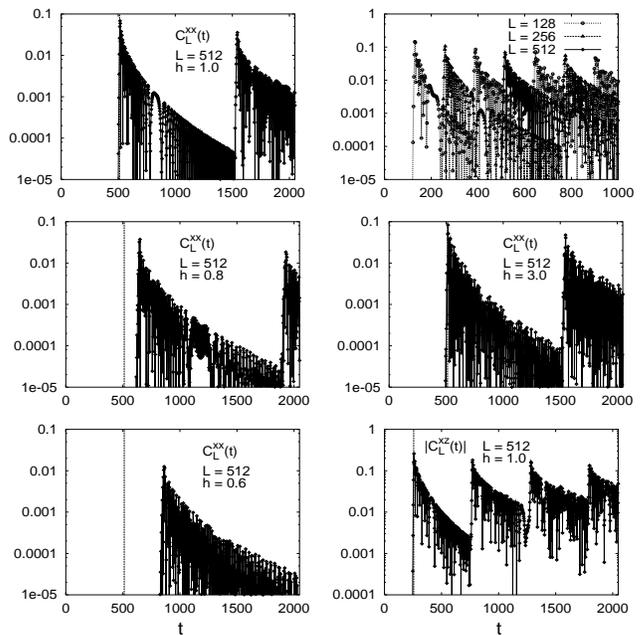}
\caption{
  Connected end-to-end correlations $\tilde{C}_L^{xx}$ (and $\tilde{C}_L^{xz}$
  for bottom left) with fixed system size $L$ and field $h$ as a function of
  the time $t$ calculated with (\ref{Cxx}) and (\ref{Cxz}).  The left column
  shows data for decreasing field strength $h=1.0$, $0.8$, $0.6$, the broken
  vertical line is at $t=512=L$.  The upper right figure shows $C_L^{xx}$ for
  different system sizes at $H=1.0$, the middle right plot shows $C_L^{xx}$ at
  $h=3.0$ and the lower right figure shows the modulus of $C_L^{xz}$ for
  $h=1$. Here the broken vertical line is at $t=256=L/2$. For the
  interpretation see text.
  \label{fig1}
}
\end{figure}

The time dependence of the {\it connected} correlations (generally defined via
\ref{obs} as $\tilde{C}_{AB}(t_1,t_2)=C_{AB}(t_1,t_2)-A(t_1)B(t_2)$)
$\tilde{C}^{x,x}_L(t)=|\langle A_1 B_L \rangle_t|^2$ shows the following
features which can be read from fig.\ \ref{fig1}: 1) They are zero for times
smaller than a time $\tau_h(L)$ which is equal to the system size $L$ for
$h\ge1$ and increases monotonically with decreasing $h$ for $h<1$. 2) At
$t=\tau_h(L)$ a jump occurs to a value that decreases algebraically with the
system size $L$:
\be
\tilde{C}^{x,x}_{\rm max}(L)
=\tilde{C}^{x,x}_L(t=\tau_h(L))\propto L^{-a}\;,
\ee
with $a=2/3$ for $h=1$ and $a=1/2$ for $h>1$. 3) For $t\ge\tau_h(L)$
the correlations decay slower than exponentially, roughly with a
stretched exponential. 4) For $t=3\tau_h(L)$ again a sudden jump
occurs as for $t=\tau_h(L)$ followed by a slightly slower oscillatory
decay. 5) This pattern is repeated for time
$t=5\tau_h(L),7\tau_h(L),\ldots$, but gets progressively smeared out
by oscillations.

These features can be interpreted as follows: the elementary (tunnel)
processes of the quantum dynamics of the Hamiltonian (\ref{hamilton})
are spin flips induced by the transverse field operator $\sigma_l^z$.
In this picture two spins can only act coherently and thus give a
contribution to the connected correlation function if the information
about such a spin flip processes reaches the two spins 
simultaneously. Feature 1 tells us that signals generated in the
center of the system travel with a speed of proportional to
$L/\tau_h(L)$ to the boundary spins and reaches both simultaneously.
At this moment $\tilde{C}^{x,x}_L(t)$ jumps to its maximum (see 2).
After this, this signal is superposed by other more incoherent signals
(see 3). However, the strongest initial signal is reflected at both
boundaries and reaches the opposite boundary spins simultaneously
again at time $t=3\tau_h(L)$ (see 4), and so on. Of course more
and more incoherent processes occur in the meantime, giving rise to
feature 5.

A similar behavior can be observed for the end-to-end correlations
when starting with the state $\ket{z}$, which is
\be
C^{x,z}_L(t)
=\sum_k \left( \langle A_1 B_k \rangle_t \langle B_L A_k \rangle_t 
- \langle A_1 A_k \rangle_t\langle B_L B_k \rangle_t \right)\;.
\label{Cxz}
\ee
The only difference to the behavior of $C^{x,x}_L(t)$ reported above is a) its
long time limit vanishes for all values of $h$ and b) $\tau_h(L)$, i.e.\ the
earliest time at which the two boundary spins are correlated, is only half as
big as in the previous case.  Obviously it is easier to generate and to
propagate spin flip signals when starting with a $z$-state.

Next we study the {\it bulk} behavior of the expectation values and
correlations involving $\sigma_l^z$ operators. We start its
non-equilibrium expectation value
\beqn
e_l^\psi(t)&=&\bra{\psi_0}\sigma^z_l(t)\ket{\psi_0}\label{el}\\
&=&\sum_k\left( \langle A_l B_k \rangle_t \langle B_l A_{i(k)}\rangle_t 
- \langle A_l A_{i(k)} \rangle_t\langle B_l B_k \rangle_t \right)\;,
\nonumber
\eeqn
with $i(k)=k,(k+1)$ for $\psi=z,(x)$. We note that the equilibrium
(i.e.\ ground state) expectation value, $e_l^0$, corresponds to the
energy-density in the two-dimensional classical Ising model and we use
this terminology also in this non-equilibrium situation. For long
times the non-equilibrium energy-density approaches a finite limit,
$\overline{e}^\psi_l$, which for a bulk spin is a) for the initial
state $\ket{\psi_0}=\ket{x}$ given by $\overline{e^x}=h/2$ for $h\le1$
and $\overline{e^x}=1/(2h)$ for $h>1$.  and b) for the initial state
$\ket{\psi_0}=\ket{z}$ by $\overline{e^z}=1/2$ for $h\le1$ and
$\overline{e^z}=1-1/2h^2$ for $h>1$. Therefore the analogue to the
specific heat $c_v \sim \partial \overline{e^z}/ \partial h$ is
discontinuous at the transition point. The relaxation of the
energy-density to its stationary value is algebraic and follows a
$\sim t^{-3/2}$ low for any value of the transverse field, $h$. At the
transition point, $h=1$, we have the analytical results in terms of
the Bessel-function, $J_{\nu}(x)$: $e_l^\psi(t)=1/2 \pm J_1(4t)/4t$,
where the + (--) sign refer to $\psi=z(x)$.

The two-spin non-equilibrium dynamical and spatial correlations
involve contributions from different processes described by the
contractions (\ref{contr}) and the corresponding formulas are
complicated, therefore they will be presented elsewhere\cite{unp}.
Here we report on the basic features of the asymptotic behavior of
correlations.  The two-time correlation function ($t_1\le t_2$), 
\be
G_l^{z,\psi}(t_1,t_2)=
\bra{\psi_0}\{\sigma_l^z(t_1)\sigma_l^z(t_2)\}_S\ket{\psi_0}\;,
\ee
is non-stationary for intermediate times, $t_1/(t_2-t_1)=O(1)$, which can be
read off from our analytical result for the connected bulk auto-correlations
at $h=1$:
%
%\beqn
%\tilde{G}_{\rm bulk}^{z,\psi}(t_1,t_2)
%=& &J_0^2(2t_2-2t_1) \label{Gzpsi}\\
%-& \frac{1}{4}\Bigl\{ & [J_2(2t_2+2t_1)+J_0(2t_2+2t_1)]\nonumber\\
% & \pm                & [J_2(2t_2-2t_1)-J_0(2t_2-2t_1)]\Bigr\}\nonumber
%\nonumber
%\eeqn
%
\be
\tilde{G}_l^{z,\psi}(t_1,t_2)
=J_0^2(2t_2-2t_1)-\frac{1}{4}[f(t_2+t_1)\pm g(t_2-t_1)]
\label{Gzpsi}
\ee
where $f(x)=J_2(2x)+J_0(2x)$, $g(x)=J_2(2x)-J_0(2x)$ and
the + (--) sign refer to $\psi=x(z)$. Thus we conclude that for
intermediate times there is {\it aging} in the $z$-component auto-correlation
function, contrary to what is reported in \cite{short}. Asymptotically we have
$\lim_{t_1 \to \infty} G_l^{z,\psi}(t_1,t_2)=(\overline{e}^\psi)^2$, and the
connected two-time correlations depends only on the time difference, e.g.\ for
$h=1$ via (\ref{Gzpsi}) $\lim_{t_1\to\infty}\tilde{G}_l^{z,\psi}(t_1,t_2) 
=J_0^2(2[t_2-t_1])-\{J_1'(2[t_2-t_1])\}^2$. For bulk
spins this stationary correlation function decays algebraically as
$\sim (t_2-t_1)^{-2}$ for any value of $h$.

\begin{figure}[t]
\epsfxsize=\columnwidth\epsfbox{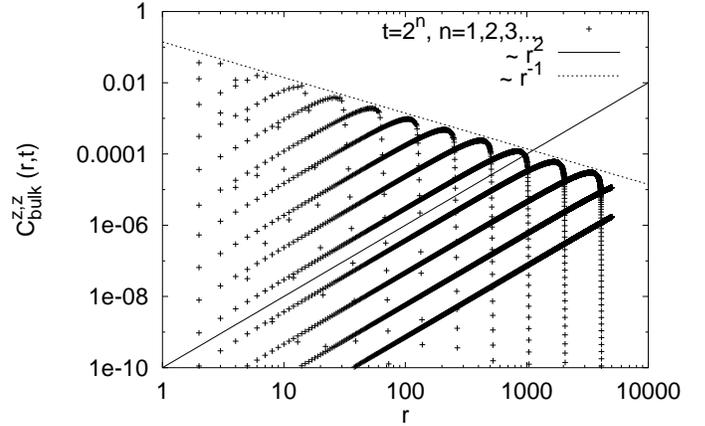}
\caption{
  Connected $\sigma_z$ correlation function
  $\tilde{C}^{z,\psi}(r,t)$ at $h=1$ given by the expression
  (\protect{\ref{corrl}}) for different times ($t=2^n$,
  $n=1,2,3,\ldots$ from left to right) in a log-log plot. The two
  straight lines indicate the initial $r^2$ dependence of
  $\tilde{C}^{z,\psi}(r,t)$ for fixed $t$ as well as the $r^{-1}$
  dependence of the maximum value at $r=2t$. 
\label{fig2}
}
\end{figure}

Next we consider the spatial equal-time correlations
\be
C^{z,\psi}(r,t)
=\bra{\psi_0}\{\sigma_{i-r/2}^z(t)\sigma_{i+r/2}^z(t)\}_S\ket{\psi_0}\;,
\ee
where $i=L/2$ in a finite system. For long times they approach the stationary
limit, $\lim_{t\to\infty} C^{z,\psi}(r,t)=(\overline{e}^\psi)^2$, the same as
for the auto-correlation function.  For the connected correlation function
$\tilde{C}^{z,\psi}(r,t)=C^{z,\psi}(r,t)-e_l^\psi(t)e_{l+r}^\psi(t)$ we can
derive an analytic expression at the transition point, $h=1$ in the limit
$L\to\infty$
\beqn
\tilde{C}^{z,\psi}(r,t)
&=& \left[{r \over 2 t} J_{2r}(4 t) \right]^2\label{corrl}\\
&-& {r^2-1 \over 4 t^2} J_{2r+1}(4 t) J_{2r-1}(4 t)\;,\nonumber
\eeqn
which is valid both for $\ket{\psi_0}=\ket{x}$ and $\ket{\psi_0}=\ket{z}$.
In fig.\ \ref{fig2} we show the $r$-dependence of $\tilde{C}^{z,z}(r,t)$ for
various times $t$. We see that for fixed time $t$ the
correlations increase proportional to $r^2$ for distances $r\le t$
to a maximum value $\tilde{C}^{z,z}_{\rm max}(t)$ at $r=2t$, which
decreases with time proportional to $t^{-1}$. For distances
larger than $r=2t$ they drop rapidly, faster than exponentially, to
zero. 

The latter two features correspond perfectly to what we observed also
for the $z$-end-to-end correlations, see eq(\ref{Cxz}): spins that are
separated by a distance $r$ can only be correlated after the first
signal from spin flip processes in between them reach simultaneously
the two spins, i.e.\ for times $t$ larger than $r/2$ (for
$h=1$ and $\ket{\psi_0}=\ket{z}$. The first feature, that correlations
for distances smaller than $2t$ are diminished only algebraically
rather than via a stretched exponential in the case of end-to-end
correlations, is new and characteristic for bulk spins. For $r\le2t$
the correlation function $\tilde{C}^{z,\psi}(r,t)$ obeys the
characteristic scaling form
\be
\tilde{C}^{z,\psi}(r,t)=t^{-1}g(r/t)
\ee
with $g(x)\propto x^2$ for $x\ll1$. The scaling parameter $r/t$
appearing in the scaling function $g(x)$ is reminiscent of the fact
that space and time scales are connected linearly at the critical point
in the transverse Ising chain since the dynamical exponent is $z=1$.
Away from the critical point we have to evaluate our expressions
\cite{unp} for $\tilde{C}^{z,\psi}(r,t)$ numerically for finite but
large ($L=512$) system sizes. Essentially we observe the same scenario as
at the critical point, the only difference being that the
general functional dependency is superposed by strong oscillations.
Moreover, starting with $\ket{\psi_0}=\ket{x}$ instead of $\ket{z}$
changes the correlations only by a constant factor.

We collect now our results for the maximum value for connected
spin-spin correlations since they decay algebraically with various new
exponents. we define ourselves to $h\ge1$ since here the time $\tau_h$
of maximum correlation is fixed, whereas for $h<1$ the value of
$\tau_h$ depends on $h$ and has to be determined numerically that
renders the precise determination of the decay exponents difficult.  We
define the ratio $\alpha=t/L$ and $\alpha_{\rm max}=\tau_h(L)/L$ and
consider equal time correlations for fixed values of $\alpha$. In the
picture of a propagating front, that separates a region in the
space-time diagram for the chain in which spins are uncorrelated from
a region in which they are correlated, one observes quasi long range
correlations {\it on the front}, the latter being defined by the ratio
$t/L=\alpha_{\rm max}$.  For distances smaller than the distance of
maximum correlation or times larger than $\tau_h$ the correlations
decay slower than exponential in time, e.g.\ algebraically for bulk
spins ($\tilde{C}^{z\psi}(t,r={\rm fixed})\sim t^{-2}$)). When
we vary both space and time with fixed ratio $t/L$ or $t/r$ we get
power laws, as long as we stay {\it behind} the front (i.e.\ 
$t\ge\tau_h$).  For $\alpha>\alpha_{\rm max}$ we observe again power
laws, but with different exponents; they are listed in table 1.

To conclude we studied a novel type of dynamically produced long range
correlations in a quantum relaxation process in a quantum spin chain.
Starting with a homogeneous initial state the quantum mechanical time
evolution according to the Schr\"odinger equation drives the system into a
stationary state, which has algebraically decaying time-dependent
autocorrelations but no critical fluctuations. However, {\it during} the
relaxation process spin-spin correlation build up upon arrival of a front of
coherent signals, which afterwards decay algebraically in the bulk. {\it On}
the front and behind it for fixed ratio of space and time scales one observes
quasi long range order.  This does {\it not} depend on any external parameter
like the transverse field. This type of algebraic correlation needs not to be
triggered by some tuning parameter and is therefore reminiscent of phenomena
in self organized criticality \cite{BTW}.  The scenario we have reported here
is a result of quantum interference and one may expect that a similar one
holds for other quantum systems, too.  At this point one should mention the
possibility of coarsening in quantum systems as for instance reported in
\cite{rica}, which is different from the scenario we have reported here.

Acknowledgment: This study has been partially performed during our
visits in Saarbr\"ucken and Budapest, respectively. F.\ I.'s work has
been supported by the Hungarian National Research Fund under grant No
OTKA TO23642, TO25139, MO28418 and by the Ministery of Education under
grant No. FKFP 0596/1999.

\begin{center}
\begin{tabular}{|l|c||c|c||c|c|}
\hline
 \multicolumn{2}{|c||}{} & \multicolumn{2}{c|}{$h=1$} & \multicolumn{2}{c|}{$h>1$} \\
  \cline{1-2} \cline{3-6}
 & $\alpha_{\rm max}$ 
 & $\alpha$=$\alpha_{\rm max}$ & $\alpha\!\!>\!\!\alpha_{\rm max}$
 & $\alpha$=$\alpha_{\rm max}$ & $\alpha\!\!>\!\!\alpha_{\rm max}$\\
\hline
\hline
$\tilde{C}_L^{xx}(t$=$\alpha L)$ & $1$   & $L^{-2/3}$ & $L^{-1}$ 
                                   & $L^{-1/2}$   & $L^{-1}$\\
$\tilde{C}_L^{xz}(t$=$\alpha L)$ & $1/2$ & $L^{-1/4}$ & $L^{-1/2}$ 
                                   & $ - $   & $ - $\\
$C_L^{z\psi}(t$=$\alpha L)$      & $1/2$ & $L^{-5/4}$ & $L^{-1}$ 
                                   & $L^{-5/8}$ & $L^{-1}$\\
$\tilde{C}^{z\psi}(t$=$\alpha r)$& $1/2$ & $r^{-4/3}$ & $r^{-1}$ 
                                   & $r^{-2/3}$ & $r^{-1}$\\
\hline
\end{tabular}
TABLE I
\end{center}


\begin{references}
\bibitem{pfaff} 
        For the equilibrium system the equal-time correlations can be 
        written in the form of a $r \times r$ determinant, for the 
        non-equilibrium problem it involves the sum of $O(r!)$
        determinants \cite{unp}.
\end{references}
\end{document}